\documentstyle[prl,preprint,aps,epsf]{revtex}

\def\addcontentsline#1#2#3{\relax}
\begin{document}
\title{\Large\bf Scattering of Conduction Electrons \\
by a Ferromagnetic Domain Wall
\footnote{Proceedings of the 4th International Symposium on Physics
of Magnetic Materials (ISPMM'98), August 1998, Sendai Japan.}}
\author{Masanori Yamanaka and Tohru Koma$^1$}
\address{Department of Applied Physics, Science University of Tokyo,
Kagurazaka, Shinjuku-ku, Tokyo 162, Japan}
\address{${}^1$Department of Physics, 
Gakushuin University, Mejiro, Toshima-ku, Tokyo 171, Japan}
\maketitle
\begin{abstract}
We study the scattering of an electron by a ferromagnetic domain wall 
of the quantum Heisenberg-Ising model (XXZ model) 
with certain boundary conditions. 
The spin of the electron interacts with the spins of the XXZ model 
by the Hund coupling. Using the exact domain wall ground states of 
the XXZ model, we analytically obtain the exact effective 
Schr\"odinger equation for conduction electrons. 
This equation coincides with a conventional phenomenological 
Schr\"odinger equation which was derived in a classical treatment of 
spins of a domain wall. 
By solving the Schr\"odinger equation numerically, 
we have calculated the transmission coefficient 
which is a function of the Hund coupling and of the anisotropy of 
the XXZ model. It turns out that  the transmission coefficient is vanishing 
in the low energy limit for the electron. 
\bigskip

\end{abstract}

\section{Introduction}
The early experiments of ferromagnetic materials \cite{REFtaylor} 
show that the resistivity increases with the increase of the number of 
the domain walls. This result seems to admit of no doubt 
because the conduction electrons 
are scattered by the domain walls \cite{REFcf}. 
An opposite result, however, was reported recently, i.e., 
the existence of a {\em negative} domain wall contribution to 
the resistivity was proposed from experimental \cite{REFhg,REFotani,REFparkin} 
and theoretical \cite{REFtatara} studies. In particular, Tatara and 
Fukuyama \cite{REFtatara} stated that the existence of impurities plays 
an important roll for the negative domain wall contribution. 
We should note that some other recent studies 
\cite{REFgregg,REFmibu,REFyn,REFlz} 
show that a domain wall contribution to resistivity is {\em positive}.
As a starting point, it is necessary to study the problem of 
the scattering of conduction electrons by a single domain wall. 
This problem was considered in the early works \cite{REFcf} 
by Cabrera and Falicov. But they treated the spins of the domain wall 
classically. 

In this paper, we study the scattering of an electron by a {\em quantum} 
domain wall which is realized as a ground state of 
the quantum Heisenberg-Ising model (XXZ model) 
with certain boundary conditions \cite{REFasw,REFgw,REFmatsui,REFkn}. 
The spin of the conduction electron interacts with the spins of the XXZ model 
by the Hund coupling. Our main results are the following two: 
(i) We analytically obtain the exact effective Schr\"odinger equation 
for the conduction electron. More precisely the spin of the conduction 
electron feels the effective magnetic field of $\tanh(x/\lambda)$, 
and the amplitude of the spin flip is proportional to $1/\cosh(x/\lambda)$. 
Here $x$ is the one-dimensional coordinate perpendicular to 
the domain wall centered at $x=0$, and $\lambda$ is the width of 
the domain wall. Our effective Schr\"odinger equation 
coincides with a conventional phenomenological Schr\"odinger equation 
by Cabrera and Falicov \cite{REFcf}. 
{From} this equation, one can immediately notice that 
the domain wall never transmits an electron 
with low energy without a spin flip of the electron. 
(ii) By solving the effective Schr\"odinger equation numerically, 
we have calculated the transmission coefficient which is a function of 
the Hund coupling and of the width of the domain wall. 
(The width is determined by the anisotropic exchange of the XXZ model.) 
It turns out that the transmission coefficient 
is vanishing in the low energy limit for the electron.  
As an example of a real system, consider a quantum wire with a single 
domain and with low density impurities. Then the transmission coefficient 
we obatined leads to the conductance by relying on the Landauer formula. 
Unfortunately it is not so easy to realize this situation in an experiment.

Throughout the present paper, we assume that the systems we consider 
are spatially homogeneous except for the $x$ direction perpendicular 
to domain walls. Having this assumption in mind, we treat the systems 
in any dimension as a one-dimensional one in this paper. 

\section{A classical domain wall}

Before discussing a quantum domain wall, we briefly review a classical 
treatment of a domain wall. The spin configuration of a classical domain wall 
(Bloch wall) is represented by a vector-valued function ${\vec M}(x)$ of 
the position $x$. In \cite{REFcf}, the authors assumed 
$\vec{M}(x)=B(-{\rm sech}{(x/\lambda)}, 0,-\tanh(x/\lambda))$, 
where $\lambda$ is the width of the domain wall, and $B$ is a real 
constant. The interaction between ${\vec M}(x)$ and the spin 
${\vec s}=(s^{(1)},s^{(2)},s^{(3)})$ 
of a conduction electron is given by the Hund coupling 
$(M^{(1)}(x)s^{(1)}+M^{(2)}s^{(2)}+\Delta M^{(3)}s^{(3)})$ with 
the anisotropy $\Delta$. 
Then the Schr\"odinger equation for the conduction electron is given by 
\begin{equation}
\left\{
\begin{array}{@{\,}l}
\displaystyle{\left[-\frac{\hbar^2}{2m}\frac{d^2 }{d x^2}
-\frac{B\Delta}{2}\tanh\left(\frac{x}{\lambda}\right)-E\right]
\psi_{\uparrow}
=\frac{B}{2}{\rm sech}\left(\frac{x}{\lambda}\right)\psi_{\downarrow}},
\nonumber\\
\quad
\nonumber\\
\displaystyle{\left[-\frac{\hbar^2}{2m} \frac{d^2}{d x^2}
+\frac{B\Delta}{2}\tanh\left(\frac{x}{\lambda}\right)-E\right]
\psi_{\downarrow}
=\frac{B}{2}{\rm sech}\left(\frac{x}{\lambda}\right)\psi_{\uparrow},}
\end{array}
\right.
\label{phenoSchro}
\end{equation}
where $\psi_{\uparrow}$ and $\psi_{\downarrow}$ are, respectively, 
the up ($\uparrow$) and the down ($\downarrow$) components of the wavefunction 
of the electron, and $m$ is the mass of the electron.
In the region far right from the domain wall ($x\sim\infty$), 
a up-spin electron has lower energy of the Hund coupling than an 
down-spin electron. In the opposite region ($x\sim-\infty$), 
up and down spins interchange their situation. If an electron goes thorough 
the domain wall without a spin flip of the electron, 
then the electron feels the potential barrier. A spin flip of the 
electron occurs only near the domain wall. 

In this treatment, the effective magnetic potential for the conduction 
electrons is assumed to be identical to the profile of the domain wall. 
Clearly the potential should be derived from a microscopic Hamiltonian 
because the spins of the conduction electrons interact 
with {\em quantum spins} of a domain wall. 
The domain wall also should be realized in the microscopic Hamiltonian. 
Thus the phenomenological Schr\"odinger equation 
by Cabrera and Falicov needs to be justified from a microscopic level.

\section{A quantum domain wall}
\label{section:quantumwall} 

In order to introduce a quantum domain wall, 
we consider the spin-1/2 Heisenberg-Ising model (XXZ model) on 
the one-dimensional lattice $[-L,L]$. The Hamiltonian 
is given by 
\begin{equation}
H_{\rm dw}= -J_H 
 \sum_{y=-L}^{L-1}
\left( S_y^{(1)} S_{y+1}^{(1)}+S_y^{(2)} S_{y+1}^{(2)}
+ \Delta_H S_y^{(3)} S_{y+1}^{(3)}\right)
+\frac{J_H}{2}\sqrt{(\Delta_H)^2-1}\left[S_{-L}^{(3)}-S_L^{(3)}\right],
\label{hamHei}
\end{equation}
where ${\vec S}_y=(S^{(1)}_y,S^{(2)}_y,S^{(3)}_y)$ is the spin-1/2 
operator at the site $y\in[-L,L]$, and $J_H,\Delta_H$ are the exchange 
integral and the anisotropy, respectively. We assume $J_H>0$ and $\Delta_H>1$. 
We apply the boundary fields to make a single domain wall as a ground state. 
The sector of the ground states \cite{REFasw,REFgw,REFmatsui,REFkn} 
of $H_{\rm dw}$ for a finite size $L$ is spanned by the set of 
the product states \cite{REFgw} 
\begin{equation}
\Psi(z)=\left|\left.\eta_{-L}(z),\eta_{-L+1}(z),\ldots,\eta_{L-1}(z),
\eta_L(z)\right\rangle\right., 
\label{equation:productstate}
\end{equation}
where $z$ is a complex number, and 
\begin{equation}
\eta_y(z)=\frac{1}{\sqrt{1+|z|^2q^{2y}}}
\left(|\uparrow\rangle_y+zq^y|\downarrow\rangle_y\right). 
\label{eta}
\end{equation}
Here $q$ is defined as $\Delta_H=(q + q^{-1})/2$ with $0<q<1$, 
and $|\uparrow\rangle$ and $|\downarrow\rangle$ are the spin 
up and down states, respectively. 
Write $z=e^{\mu+i\phi}$ with real numbers $\mu,\phi$. 
Then $\mu$ specifies the position of the domain 
wall and the angle $\phi$ is a quantum mechanical phase corresponding to 
the degrees of freedom of the rotation about the third axis of the spin. 
For simplicity we set $z=1$. Then we have, from (\ref{eta}), 
\begin{equation}
\eta_y(z)\sim \cases{|\uparrow\rangle_y & for $y\sim \infty$\cr
                     |\downarrow\rangle_y & for $y\sim-\infty$\cr} 
\end{equation}
in the thermodynamic limit $L\to+\infty$, and from 
(\ref{equation:productstate}),
\begin{equation}
\langle\Psi(z), S^{(3)}_y \Psi(z)\rangle=\frac{1}{2}
\tanh\left(\frac{x}{\lambda}\right) 
\ \ \mbox{with} \ \lambda=-\frac{a}{\log{q}},
\label{eq:elementz}
\end{equation}
where $a$ is the lattice constant, and we have used $x=ay$. 
Thus the expectation about the {\em quantum} domain 
wall ground state $\Psi(z)$ gives the profile of the classical domain wall. 

\section{A Heisenberg-Kondo model}

Now we consider the problem of the scattering of a conduction electron 
by a single quantum domain wall.
To treat this problem,
we introduce a Heisenberg-Kondo model. 
See Fig.~\ref{FIGUREmodel} for the schematic structure. 
The Hamiltonian is given by 
\begin{equation}
H=H_{\rm el}+H_{\rm dw}+H_{\rm el-dw}
\label{eq:hamiltonian}
\end{equation}
with the kinetic energy of the conduction electron 
\begin{equation}
H_{\rm el}=-\frac{\hbar^2}{2m} \frac{d^2}{dx^2}
\end{equation}
and the Hund coupling 
\begin{eqnarray}
H_{\rm el-dw}
=-J_K\sum_{y=-L}^{L}{\cal P}
\left[
{\vec s}\cdot{\vec S}_y
+(\Delta_K-1)s^{(3)} S_{y}^{(3)}\right]{\cal P}
\times\ \chi_y(x),
\end{eqnarray}
between the spin ${\vec s}$ of the electron 
and the spins ${\vec S}_y$ of the domain wall of $H_{\rm dw}$ of 
(\ref{hamHei}). Here $J_K,\Delta_K$ are real parameters, 
${\cal P}$ is the projection operator onto 
the sector spanned by the ground states of $H_{\rm dw}$, and 
$\chi_y(x)$ is a characteristic function defined by 
\begin{eqnarray}
\chi_y(x) = \left\{\begin{array}{ll}
1, & \mbox{if}\ x \in a [y, y+1]\\
0, & \mbox{otherwise.}
\end{array}
\right.
\end{eqnarray}
By introducing the projection ${\cal P}$, we have neglected the excitations 
above the domain wall ground states of the XXZ model. In pure one dimension, 
this approximation is justified because of the energy gap above those ground 
states \cite{REFkn}. But, in higher dimensions there exist gapless 
excitations above the domain wall ground states \cite{REFmatsui,REFkn}. 
The treatment of the excitations is left for future studies 
\cite{REFyamanakakoma}. 

Using the domain wall ground state $\Psi(z)$ of (\ref{equation:productstate}), 
a single-electron ``eigenstate" can be written as 
\begin{eqnarray}
\Phi(z) &=&\left(
\psi_{\uparrow}~\vert \uparrow \rangle +
\psi_{\downarrow}~\vert \downarrow \rangle \right)\otimes\Psi(z),
\label{eq:oneparticle}
\end{eqnarray}
where $\psi_{\sigma}(x)$ is the $\sigma$ component of the wave function 
of the conduction electron. 
Then the effective Schr\"odinger equation for the conduction electron 
is given by 
\begin{equation} 
\langle\Psi(z), H\Phi(z)\rangle=E\langle\Psi(z),\Phi(z)\rangle, 
\label{eq:schroedinger}
\end{equation} 
where we have used the single-electron Schr\"odinger equation 
$H\Phi=E\Phi$ for the Hamiltonian $H$ of (\ref{eq:hamiltonian}) 
and with the energy eigenvalue $E$. 
Using (\ref{eq:elementz}) and 
$2\langle\Psi(z), S_y^{(\pm)}\Psi(z)\rangle={\rm sech}(x/\lambda)$, 
the equation (\ref{eq:schroedinger}) is written as 
\begin{equation}
\left\{
\begin{array}{@{\,}l}
\displaystyle{\left[-\frac{\hbar^2}{2m}\frac{d^2}{dx^2}
-\frac{J_K \Delta_K}{4}\tanh\left(\frac{x}{\lambda}\right)-E\right]
\psi_{\uparrow}
=\frac{J_K}{4}{\rm sech}\left(\frac{x}{\lambda}\right)\psi_{\downarrow},}
\nonumber\\
\quad
\nonumber\\
\displaystyle{\left[
-\frac{\hbar^2}{2m}\frac{d^2}{dx^2}
+\frac{J_K \Delta_K}{4} \tanh\left(\frac{x}{\lambda}\right)-E\right]
\psi_{\downarrow}
=\frac{J_K}{4}{\rm sech}\left(\frac{x}{\lambda}\right)\psi_{\uparrow}.}
\end{array}\right.
\label{eq:schrodingereq}
\end{equation}
This coincides with the phenomenological equation (\ref{phenoSchro}) 
by Cabrera and Falicov. Although the vector (\ref{eq:oneparticle}) 
is {\em not} an eigenvector of the Hamiltonian $H$ of (\ref{eq:hamiltonian}) 
in a usual sense, we can justify the above treatment 
for the present system in any dimension \cite{REFyamanakakoma}. 
Suppose that an $\uparrow$ electron comes from the right side of 
the domain wall. 
Then the spin of the electron feels the potential barrier 
$-(J_K\Delta_K/4)\tanh(x/\lambda)$ 
and flips with the transition amplitude $-(J_K/4){\rm sech}(x/\lambda)$ 
near the domain wall. 
Since the height of the potential wall is order of $J_K$, 
a conduction electron with low energy $(E<J_K\Delta_K/4)$ 
never goes to the other side without a spin flip.

\section{Numerical results and discussion}

We solved the Schr\"odinger equation (\ref{eq:schrodingereq}) numerically 
and obtained the transmission coefficient 
as a function of the energy $E$ of the conduction electron 
and of the width $\lambda$ of the domain wall. 
The results for the transmission coefficients with a spin flip are shown 
in Fig.~\ref{FIGUREtransmission}. 
The remarkable feature is that the transmission coefficients are vanishing 
in the low energy limit for the conduction electron. 
As we expected, our results indicate that there appears no negative domain 
wall contribution to resistivity without introducing a new mechanism 
such as an effective interaction between domain walls and impurities. 
Actually such mechanisms were proposed by some authors \cite{REFtatara,REFlz} 
for explaining the recent experiments 
\cite{REFhg,REFotani,REFparkin,REFgregg,REFmibu}. 
However, it is very difficult to take account of the effect of impurities in 
a mathematically rigorous manner as in the present paper. 
\bigskip

\noindent
{\bf Acknowledgements\ \ }The authors are grateful 
to Koh~Mibu, Taro~Nagahama, Naoto~Nagaosa, Yoshiko~Nakamura,
Yoshichika~Otani, Teruya~Shinjo, and Gen~Tatara for useful discussions. 
A part of the numerical calculations was performed by using facilities 
of the supercomputing center in ISSP, University of Tokyo. 



\begin{figure}[b]
\bigskip
\caption{
A schematic description of the Heisenberg-Kondo model. 
The open circles denote the spins of the XXZ model. 
The conduction electron moves along the dotted line. 
The bonds of the solid lines denote the ferromagnetic exchange interactions 
of the XXZ model. The broken lines denote the Hund couplings 
between the spins of the XXZ model and the spin of the conduction electron. 
}
\label{FIGUREmodel}
\end{figure}

\vspace{10mm}

\begin{figure}[b]
\bigskip
\caption{
The transmission coefficients as a function of the energy $E$ of 
the conduction electron in the unit of $J_K \Delta_K$. 
}
\label{FIGUREtransmission}
\end{figure}


\begin{thebibliography}{99}
\bibitem{REFtaylor}
G.R.~Taylor, S.~Isin, and R.V.~Coleman,
Phys. Rev. {\bf 165}, 621 (1968).

\bibitem{REFcf}
G.G.~Cabrera and L.M.~Falicov,
Status Solidi (b) {\bf 61}, 539 (1974);
ibid {\bf 62} 217 (1974).


\bibitem{REFhg}
K.~Hong and N.~Giordano
Phys. Rev. B{\bf 51}, 9855 (1995);
J. Magn. Magn. Mat. {\bf 151}, 396 (1995);
J. Phys. Condens. Matt. 
{\bf 8}, L301 (1996);
ibid {\bf 10}, L401 (1998).

\bibitem{REFotani}
Y.~Otani, K.~Fukamichi, O.~Kitakami, Y.~Shimada,
B.~Pannetier, J.-P.~Nozieres, T.~Matsuda, and A.~Tonomura,
Proceedings of MRS Spring meeting, (San Francisco)
{\bf 475}, 215 (1998).

\bibitem{REFparkin}
U.~Ruediger, J.Yu.~Zhang, A.D.~Kent, and S.S.P.~Parkin,
{\it Phys. Rev. Lett.},
{\bf 80}, 5639 (1998).

\bibitem{REFtatara}
G.~Tatara and H.~Fukuyama, 
Phys. Rev. Lett. {\bf 78}, 3773 (1997).


\bibitem{REFgregg}
J.F.~Gregg, W.~Allen, K.~Ounadjela, M.~Viret, M.~Hehn, S.M.~Thompson,
and J.M.D.Coey, Phys. Rev. Lett. {\bf 77}, 1580 (1996).

\bibitem{REFmibu}
K.~Mibu, T.~Nagahama, and T.~Shinjo, J. Magn.  Magn. Mater.
{\bf 163} 75 (1996);
K.~Mibu, T.~Nagahama, and T.~Ono, T.~Shinjo, unpublished.

\bibitem{REFyn}
M.~Yamanaka and N.~Nagaosa, 
J. Phys. Soc. Jpn. {\bf{65}}, 3088 (1996).

\bibitem{REFlz}
P.M.~Levy and S.~Zhang, 
Phys. Rev. Lett. {\bf 79}, 5110 (1997)


\bibitem{REFasw}
F.C.~Alcaraz, S.R.~Salinas, and W.F.~Wreszinski, 
Phys. Rev. Lett. {\bf 75}, 930 (1993).

\bibitem{REFgw}
C.-T.~Gottstein and R.F.~Werner, cond-mat/9501123.

\bibitem{REFmatsui}
T.~Matsui, Lett. Math. Phys. {\bf 37}, 397 (1996); ibid {\bf 42}, 
229 (1997).

\bibitem{REFkn}
T.~Koma and B.~Nachtergaele, Lett. Math. Phys. {\bf 40}, 1 (1997);
RIMS Kokyuroku (Kyoto University, Kyoto, Japan) No.~{\bf 1035}, 133 (1998);
{\it Adv. Theor. Math. Phys.} {\bf 2}, (1998), to appear.

\bibitem{REFyamanakakoma}
M.~Yamanaka and T.~Koma, in preparation.

\end{thebibliography}
\end{document}